\documentclass[iop]{emulateapj}

\usepackage[]{hyperref,epstopdf,amsmath,xcolor}
\usepackage[]{empheq}
\usepackage[table]{}

\slugcomment{Draft \today}

\shorttitle{The effect of underestimated field strengths on spin-down}
\shortauthors{V. See et al.}

\begin{document}

\title{How much do underestimated field strengths from Zeeman-Doppler imaging affect spin-down torque estimates?}

\author{Victor See$^{1}$, Lisa Lehmann$^{2}$, Sean P. Matt$^{1}$, Adam J. Finley$^{1}$}
\affil{$^{1}$University of Exeter, Deparment of Physics \& Astronomy, Stocker Road, Devon, Exeter, EX4 4QL, UK\\
$^{2}$SUPA, School of Physics and Astronomy, University of St Andrews, North Haugh, St Andrews KY16 9SS, UK
}
\email{*w.see@exeter.ac.uk}

\begin{abstract}
Numerous attempts to estimate the rate at which low-mass stars lose angular momentum over their lifetimes exist in the literature. One approach is to use magnetic maps derived from Zeeman-Doppler imaging (ZDI) in conjunction with so-called ``braking laws''. The use of ZDI maps has advantages over other methods because it allows information about the magnetic field geometry to be incorporated into the estimate. However, ZDI is known to underestimate photospheric field strengths due to flux cancellation effects. Recently, \citet{Lehmann2018ZDI} conducted synthetic ZDI reconstructions on a set of flux transport simulations to help quantify the amount by which ZDI underestimates the field strengths of relatively slowly rotating and weak activity solar-like stars. In this paper, we evaluate how underestimated angular momentum-loss rate estimates based on ZDI maps may be. We find that they are relatively accurate for stars with strong magnetic fields but may be underestimated by a factor of up to $\sim$10 for stars with weak magnetic fields. Additionally, we re-evaluate our previous work that used ZDI maps to study the relative contributions of different magnetic field modes to angular momentum-loss. We previously found that the dipole component dominates spin-down for most low-mass stars. This conclusion still holds true even in light of the work of \citet{Lehmann2018ZDI}.

\end{abstract}


\section{Introduction}
\label{sec:Intro}
Stellar winds carry angular momentum away from low-mass stars ($M_{\star} \lesssim 1.3 M_{\odot}$) causing them to spin-down over their main-sequence lifetime \citep{Bouvier2014}. Using magnetohydrodynamic (MHD) simulations, numerous authors have attempted to quantify the rate at which angular-momentum is lost as a function of stellar parameters \citep{Matt2012,Reville2015,Garraffo2016,Pantolmos2017,Finley2017,Finley2018}. These studies typically express the angular momentum-loss rate as a semi-analytic expression known as a braking law. A key input of these braking laws is the stellar magnetic field strength and geometry. 

Determining the magnetic properties of low-mass stars is challenging. One method is to use Zeeman-Doppler imaging (ZDI) which is a tomographic imaging technique \citep{Semel1989,Brown1991,Donati1997,Donati2006}. The large-scale magnetic field strength and geometry can be recovered from a time-series of spectropolarimetric observations and the resulting map is usually represented by a spherical harmonic decomposition. The ability for ZDI to reconstruct the field as a superposition of individual spherical harmonic modes is key for the study of spin-down torques and many studies have used ZDI maps to estimate the rate at which low-mass stars lose angular momentum \citep{Vidotto2014Torque,Vidotto2015,AlvaradoGomez2016,Reville2016,Nicholson2016,See2017,See2018}.

\begin{figure*}
	\begin{center}
	\includegraphics[trim=0cm 1cm 1cm 0cm,width=0.8\textwidth]{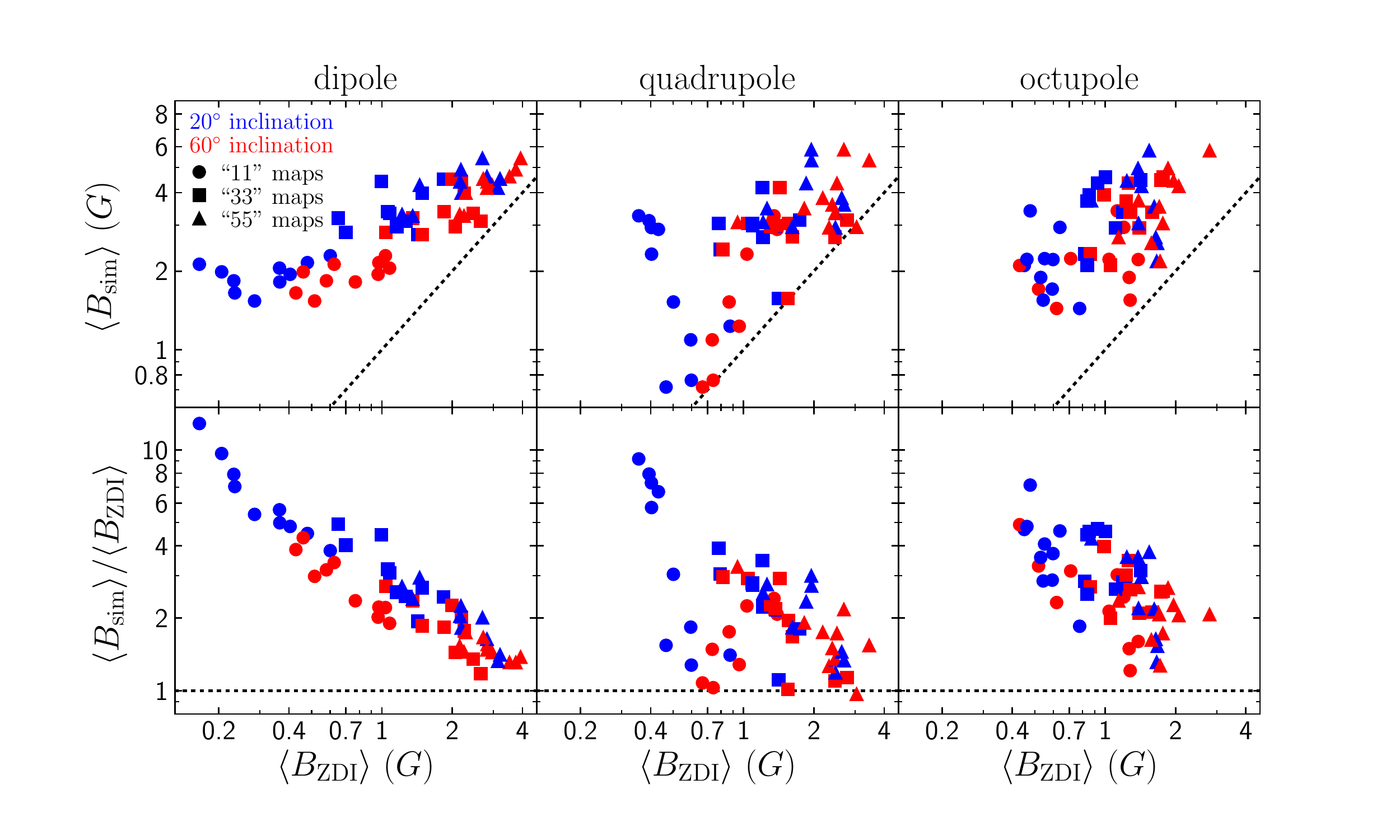}
	\end{center}
	\caption{Top: average unsigned flux from the simulations against the average unsigned flux from the ZDI reconstructions. Bottom: ratio of the average unsigned flux from the simulation to the average unsigned flux from the ZDI reconstruction against the average unsigned flux from the ZDI reconstructions. Each quantity is shown for the dipole (left), quadrupole (middle) and octupole (right) components. Blue and red points correspond to an assumed inclination of 20$^{\circ}$ and 60$^{\circ}$ respectively in the ZDI reconstruction. Circular, square and triangular points correspond to simulations with solar flux emergence rate and differential rotation (labelled ``11"), simulations with three times these quantities (labelled ``33'') and five times these quantities (labelled ``55'') respectively. Dotted lines indicate where the average unsigned flux from the simulations and ZDI reconstructions are equal.}
	\label{fig:BComp}
\end{figure*}

One drawback of ZDI is that it typically just uses circularly polarised light which is only sensitive to the large-scale magnetic field components (although some studies do incorporate unpolarised and linearly polarised light, e.g. \citet{Rosen2015}). As such, ZDI is not capable of completely recovering magnetic fields organised on small scales. ZDI generally recovers less than $\sim$10\% of the surface averaged unsigned magnetic flux\footnote{In this work, the averaged unsigned magnetic flux refers to the absolute value of the magnetic field strength, considering all 3 vector components of the field, averaged over the stellar surface.} and up to $\sim$25\% in the best cases \citep{Reiners2009,Morin2010,Kochukhov2019,See2019}. However, estimating spin-down torques using ZDI maps should not be greatly affected by the missing small-scale flux as the torque is dependent mostly on the large-scale field components \citep{Finley2018}. Indeed, a number of studies have shown that the open magnetic flux, and hence angular momentum-loss, is dominated by the dipolar component of the magnetic field for the majority of stars \citep{See2017,See2018} as well as the Sun \citep{Jardine2017}.

Recently, \citet{Lehmann2018ZDI} conducted a study into the robustness of ZDI \citep[specifically, the ZDI implementation described in][]{Hussain2016}. These authors conducted synthetic ZDI observations of the high resolution flux transport simulations performed by \citet{Gibb2016}. The types of stars studied by \citet{Gibb2016} have relatively weak activity and are slow rotators well within the unsaturated regime. \citet{Lehmann2018ZDI} then compared the reconstructed ZDI maps to the known photospheric magnetic field of the flux transport simulations. In order for this to be a fair comparison, they restricted the magnetograms from the flux transport simulations to the large-scale component by only considering spherical harmonic modes equal to or smaller than $l=7$. \citet{Lehmann2018ZDI} found that ZDI did a reasonable job of reconstructing the major features of the large-scale magnetic field of the flux transport simulations, showing good agreement up to $l\sim3$, but that the ZDI maps contained roughly an order of magnitude less magnetic energy in the low order field modes. This is the first time that a study has quantified the amount of magnetic flux that ZDI misses at low order spherical harmonic field modes and has important implications for spin-down models.

In this work, we study the amount by which torques estimated using ZDI maps may be underestimated using the results of \citet{Lehmann2018ZDI}. We also determine if the result that angular momentum-loss is dominated by the magnetic dipole \citep{See2019Dip} is still robust in light of the results of \citet{Lehmann2018ZDI}. In section \ref{sec:ZDIvsSim}, we give an overview of the work conducted by \citet{Lehmann2018ZDI} and compare how effectively ZDI can recover the the dipole, quadrupole and octupole field modes. In section \ref{sec:Torques}, we compare spin-down torques estimated from the flux transport simulations compared to the ZDI reconstructions. We also compare critical mass-loss rates, $\dot{M}_{\rm crit}$, calculated using the flux transport simulations and the ZDI reconstructions. This is a quantity introduced by \citet{See2019Dip} to help determine if angular momentum-loss is dominated by the dipole magnetic field. Finally, we present our conclusions in section \ref{sec:Conclusions}.

\section{ZDI reconstructions of flux transport simulations}
\label{sec:ZDIvsSim}
In this section, we briefly summarise the study of \citet{Lehmann2018ZDI} and highlight the results most relevant to our work. These authors tested the capabilities of ZDI by conducting synthetic ZDI observations of high resolution magnetic maps from the theoretical models of \citet{Gibb2016}. The theoretical models consisted of a set of flux transport simulations coupled to a magnetofrictional coronal evolution model.  The flux emergence rate and differential rotation were varied across the different simulations to determine the effect on the coronal magnetic field. Each simulation was run until it reached a quasi-steady state, i.e. although the coronal magnetic field is constantly adjusting in response to the flux emerging through the photosphere, the value of global properties, such as surface flux, only vary around some mean value.

\begin{table*}
\caption{The dipole, quadrupole and octupole field strengths from the flux transport models of \citet{Gibb2016}. Each map can be identified by its simulation parameters ``11'', ``33'' or ``55'' (see text) and an arbitrarily assigned map number. For each simulated map, the dipole, quadrupole and octupole field strengths from the associated ZDI maps are listed along with the assumed inclination. The torque ratio (section \ref{subsec:TorqueRatio}) and critical mass-loss ratio (section \ref{subsec:MDotCrit}) for each combination of simulated map and ZDI reconstruction are also shown.}
\label{tab:vals}
\center
\begin{tabular}{cccccccccccc}
\hline																					
Sim	&	Map	&	$\langle B_{\rm sim,d}\rangle$	&	$\langle B_{\rm sim,q}\rangle$	&	$\langle B_{\rm sim,o}\rangle$	&	ZDI	&	$\langle B_{\rm ZDI,d}\rangle$	&	$\langle B_{\rm ZDI,q}\rangle$	&	$\langle B_{\rm ZDI,o}\rangle$	&	\multicolumn{2}{c}{$T_{\rm sim}/T_{\rm ZDI}$}	& 	$\dot{M}_{\rm crit,sim}/\dot{M}_{\rm crit,ZDI}$	\\
label	&	number	&	(G)	&	(G)	&	(G)	&	inc	&	(G)	&	(G)	&	(G)	&	min	&	max	& \\
\hline																					
11	&	1	&	2.29	&	1.23	&	1.44	&	20	&	0.601	&	0.875	&	0.777	&	1.32	&	3.41	&	54.6	\\
	&		&		&		&		&	60	&	1.04	&	0.958	&	0.619	&	1.25	&	2.07	&	9.26	\\
11	&	2	&	2.16	&	0.763	&	1.71	&	20	&	0.48	&	0.597	&	0.593	&	1.43	&	3.96	&	74.9	\\
	&		&		&		&		&	60	&	0.97	&	0.741	&	0.518	&	1.29	&	2.08	&	9.28	\\
11	&	3	&	2.05	&	0.718	&	2.1	&	20	&	0.364	&	0.466	&	0.45	&	1.59	&	4.88	&	109	\\
	&		&		&		&		&	60	&	1.08	&	0.667	&	0.43	&	1.28	&	1.8	&	4.72	\\
11	&	4	&	1.94	&	1.09	&	2.23	&	20	&	0.404	&	0.594	&	0.55	&	1.53	&	4.21	&	84.1	\\
	&		&		&		&		&	60	&	0.963	&	0.734	&	0.711	&	1.31	&	1.9	&	5.74	\\
11	&	5	&	2.12	&	1.52	&	2.21	&	20	&	0.165	&	0.5	&	0.597	&	1.71	&	10.4	&	1830	\\
	&		&		&		&		&	60	&	0.624	&	0.867	&	1.38	&	1.28	&	3.07	&	29.4	\\
11	&	6	&	1.99	&	2.32	&	2.94	&	20	&	0.206	&	0.403	&	0.639	&	1.84	&	7.98	&	224	\\
	&		&		&		&		&	60	&	0.46	&	1.03	&	1.2	&	1.41	&	3.82	&	58	\\
11	&	7	&	1.54	&	3.12	&	3.4	&	20	&	0.285	&	0.394	&	0.477	&	1.97	&	4.68	&	14.7	\\
	&		&		&		&		&	60	&	0.515	&	1.33	&	1.13	&	1.41	&	2.72	&	14.3	\\
11	&	8	&	1.65	&	3.26	&	2.22	&	20	&	0.234	&	0.356	&	0.462	&	1.95	&	5.97	&	30.9	\\
	&		&		&		&		&	60	&	0.428	&	1.35	&	1.04	&	1.38	&	3.44	&	38	\\
11	&	9	&	1.82	&	2.89	&	1.89	&	20	&	0.365	&	0.432	&	0.529	&	1.75	&	4.35	&	15.3	\\
	&		&		&		&		&	60	&	0.769	&	1.39	&	1.27	&	1.26	&	2.2	&	7.01	\\
11	&	10	&	1.84	&	2.94	&	1.55	&	20	&	0.232	&	0.403	&	0.542	&	1.79	&	6.65	&	72.2	\\
	&		&		&		&		&	60	&	0.578	&	1.33	&	1.28	&	1.27	&	2.88	&	19.8	\\
33	&	1	&	2.82	&	2.42	&	2.33	&	20	&	0.7	&	0.792	&	0.818	&	1.51	&	3.58	&	23.8	\\
	&		&		&		&		&	60	&	1.04	&	0.817	&	0.862	&	1.43	&	2.49	&	6.56	\\
33	&	2	&	3.19	&	3.05	&	3.92	&	20	&	0.65	&	0.782	&	0.855	&	1.68	&	4.3	&	33.8	\\
	&		&		&		&		&	60	&	1.35	&	1.04	&	0.987	&	1.47	&	2.2	&	4.23	\\
33	&	3	&	4.42	&	2.7	&	3.37	&	20	&	0.996	&	1.21	&	1.22	&	1.48	&	3.91	&	48.3	\\
	&		&		&		&		&	60	&	2.19	&	2.46	&	1.59	&	1.2	&	1.9	&	8.86	\\
33	&	4	&	3.33	&	2.71	&	3.36	&	20	&	1.08	&	1.21	&	1.19	&	1.41	&	2.81	&	14.8	\\
	&		&		&		&		&	60	&	2.46	&	1.62	&	1.28	&	1.22	&	1.32	&	1.42	\\
33	&	5	&	3.11	&	2.98	&	2.93	&	20	&	1.26	&	1.37	&	1.11	&	1.36	&	2.28	&	7.26	\\
	&		&		&		&		&	60	&	2.65	&	1.31	&	1.4	&	1.16	&	1.26	&	0.646	\\
33	&	6	&	2.97	&	3.04	&	4.58	&	20	&	1.16	&	1.09	&	1	&	1.51	&	2.37	&	5.82	\\
	&		&		&		&		&	60	&	2.06	&	1.55	&	1.77	&	1.27	&	1.4	&	1.38	\\
33	&	7	&	2.76	&	4.18	&	3.71	&	20	&	1.42	&	1.2	&	0.836	&	1.48	&	1.83	&	1.58	\\
	&		&		&		&		&	60	&	1.49	&	1.43	&	1.23	&	1.39	&	1.77	&	1.72	\\
33	&	8	&	3.38	&	2.99	&	2.11	&	20	&	1.06	&	1.09	&	0.836	&	1.44	&	2.9	&	12.6	\\
	&		&		&		&		&	60	&	1.84	&	1.37	&	1.05	&	1.27	&	1.74	&	2.68	\\
33	&	9	&	3.98	&	1.57	&	4.35	&	20	&	1.49	&	1.41	&	0.926	&	1.39	&	2.46	&	14.9	\\
	&		&		&		&		&	60	&	2.25	&	1.55	&	1.25	&	1.23	&	1.69	&	4.35	\\
33	&	10	&	4.5	&	3.14	&	4.47	&	20	&	1.83	&	1.74	&	1.42	&	1.36	&	2.27	&	8.85	\\
	&		&		&		&		&	60	&	1.99	&	2.77	&	1.73	&	1.24	&	2.11	&	13.4	\\
55	&	1	&	3.29	&	3.47	&	4.95	&	20	&	1.22	&	1.26	&	1.38	&	1.47	&	2.49	&	7.09	\\
	&		&		&		&		&	60	&	2.15	&	1.82	&	1.86	&	1.28	&	1.48	&	1.73	\\
55	&	2	&	3.25	&	3.07	&	3.72	&	20	&	1.35	&	1.21	&	0.872	&	1.45	&	2.23	&	5.36	\\
	&		&		&		&		&	60	&	2.25	&	0.942	&	1.39	&	1.31	&	1.44	&	0.862	\\
55	&	3	&	4.27	&	2.95	&	4.43	&	20	&	1.45	&	1.62	&	1.24	&	1.41	&	2.69	&	16.3	\\
	&		&		&		&		&	60	&	2.96	&	3.05	&	1.96	&	1.1	&	1.4	&	3.47	\\
55	&	4	&	3.98	&	4.33	&	4.23	&	20	&	2.19	&	1.85	&	1.43	&	1.33	&	1.73	&	2.35	\\
	&		&		&		&		&	60	&	2.29	&	2.51	&	2.06	&	1.23	&	1.66	&	3.08	\\
55	&	5	&	4.4	&	5.31	&	5.79	&	20	&	2.16	&	1.95	&	1.54	&	1.42	&	1.92	&	2.74	\\
	&		&		&		&		&	60	&	2.83	&	3.45	&	2.8	&	1.2	&	1.5	&	2.46	\\
55	&	6	&	4.89	&	5.84	&	3.05	&	20	&	2.18	&	1.95	&	1.39	&	1.38	&	2.1	&	3.34	\\
	&		&		&		&		&	60	&	3.74	&	2.69	&	1.76	&	1.2	&	1.32	&	0.859	\\
55	&	7	&	5.41	&	2.93	&	2.18	&	20	&	2.69	&	2.47	&	1.66	&	1.16	&	1.89	&	7.45	\\
	&		&		&		&		&	60	&	3.92	&	2.32	&	1.72	&	1.1	&	1.34	&	2.08	\\
55	&	8	&	4.6	&	3.33	&	2.69	&	20	&	2.82	&	2.48	&	1.65	&	1.16	&	1.57	&	3.42	\\
	&		&		&		&		&	60	&	3.52	&	2.47	&	1.14	&	1.15	&	1.28	&	1.65	\\
55	&	9	&	4.16	&	3.81	&	2.56	&	20	&	3.14	&	2.63	&	1.67	&	1.13	&	1.29	&	1.57	\\
	&		&		&		&		&	60	&	2.82	&	2.18	&	1.58	&	1.18	&	1.43	&	1.76	\\
55	&	10	&	4.51	&	3.59	&	3.53	&	20	&	3.2	&	2.69	&	1.62	&	1.16	&	1.37	&	2.13	\\
	&		&		&		&		&	60	&	2.71	&	2.4	&	1.7	&	1.2	&	1.59	&	3.16	\\
\hline
\end{tabular}
\end{table*}

\citet{Lehmann2018ZDI} used three different simulations from \citet{Gibb2016}; one with a solar flux emergence rate and solar level of differentiate rotation, one with three times each of these parameters and one with five times each of these parameters. These 3 simulations are labelled ``11'', ``33'' and ``55'' throughout this work. These parameters correspond to relatively weak activity and slowly rotating solar-like stars. Ten snapshots of the photospheric magnetic field were utilised from each simulation resulting in 30 high resolution magnetic field maps. For each of these maps, synthetic disc-integrated Stokes \textit{I} and \textit{V} profiles were modelled with Gaussian noise injected. The Stokes \textit{I} and \textit{V} profiles were modelled with the assumption that the star is viewed at inclinations of either $20^{\circ}$ or $60^{\circ}$ with $v\sin i$ values ranging from 0.64$\rm kms^{-1}$ to 2.58$\rm kms^{-1}$ . ZDI can then be performed using these modelled Stokes profiles resulting in a total of 60 ZDI maps. Since the input magnetic map used for each ZDI reconstruction is exactly known, a direct comparison between the two can be made. For these slowly rotating, solar-like stars, \citet{Lehmann2018ZDI} found that large-scale magnetic field geometry is reproduced reasonably well up to a spherical harmonic mode of $l\sim 3$. However, they found that the reconstructed magnetic energy can be significantly smaller, even when only comparing the lowest spherical harmonic modes. 

In the context of angular momentum-loss, the lowest order field modes are the most important and so we will focus on the poloidal dipole, quadrupole and octupole components in this work which we denote as $\langle B_{\rm dip}\rangle$, $\langle B_{\rm quad}\rangle$ \& $\langle B_{\rm oct}\rangle$. We calculate these values for each input map from the flux transport simulations and for each output ZDI map. Formally, these values correspond to the surface averaged unsigned flux when considering only the $\alpha_{\rm lm}$ and $\beta_{\rm lm}$ coefficients, i.e. $\gamma_{\rm lm}=0$, and either the $l=1$, $l=2$ or $l=3$ mode (see section 2.2 of \citet{Lehmann2018ZDI} for more details on the spherical harmonic representation of the maps). These field strengths are listed in table \ref{tab:vals}. In the top row of fig. \ref{fig:BComp}, we show the field strengths from the simulations against the reconstructed ZDI field strengths for the dipole, quadrupole and octupole components. The points are colour coded by the assumed inclination in the ZDI reconstruction and the symbol shapes correspond to the different parameters used for the flux transport simulations. The dotted lines indicate where $\langle B_{\rm sim}\rangle = \langle B_{\rm ZDI}\rangle$. It is clear that ZDI, in the majority of cases, does not recover all the magnetic flux originally present in the input maps. In the bottom row of fig. \ref{fig:BComp}, we show  $\langle B_{\rm sim}\rangle / \langle B_{\rm ZDI}\rangle$ against $\langle B_{\rm ZDI}\rangle$. These panels show the same information as the top row of fig. \ref{fig:BComp} but the factor by which the field strengths are underestimated is clearer to see. The general trend is that ZDI recovers a higher fraction of the magnetic field for maps with stronger magnetic fields. The maps with the weakest fields can be underestimated by up to a factor of $\sim$ 10. 

\section{Spin-down torques}
\label{sec:Torques}
\subsection{Finley \& Matt (2018) braking law}
\label{subsec:Torque}
To estimate the angular momentum-loss rate of our simulated stars, we will use the braking law of \citet[][henceforth, F18]{Finley2018}. It is given by

\begin{equation}
	T = \dot{M}\Omega_{\star}\langle R_{\rm A}\rangle^2
	\label{eq:Torque}
\end{equation}
where $T$ is the angular momentum loss-rate or spin-down torque, $\dot{M}$ is the mass-loss rate, $\Omega_{\star}=2\pi/P_{\rm rot}$ is the angular frequency, $P_{\rm rot}$ is the rotation period and $\langle R_{\rm A}\rangle$ is the average Alfv\'{e}n radius given by

\begin{subequations}
\label{eq:BrakingLaw}
    \begin{empheq}[left={\langle R_{\rm A}\rangle/r_{\star}={\rm max}\empheqlbrace\,}]{align}
      & K_{\rm d}\left[\mathcal{R}_{\rm d}^2\Upsilon\right]^{m_{\rm d}}
      \label{eq:BrakingLawDip} \\
      & K_{\rm q}\left[\left(\mathcal{R}_{\rm d}+\mathcal{R}_{\rm q}\right)^2\Upsilon\right]^{m_{\rm q}}
      \label{eq:BrakingLawQuad} \\
      & K_{\rm o}\left[\left(\mathcal{R}_{\rm d}+\mathcal{R}_{\rm q}+\mathcal{R}_{\rm o}\right)^2\Upsilon\right]^{m_{\rm o}}.
      \label{eq:BrakingLawOct}
    \end{empheq}
\end{subequations}
Here, $r_{\star}$ is the stellar radius, $\Upsilon = \frac{B_{\star}^2 r_{\star}^2}{\dot{M}v_{\rm esc}}$ is the wind magnetisation and $v_{\rm esc}$ is the escape velocity of the star. $\mathcal{R}_{\rm d} = B_{\rm d}/B_{\star}$, $\mathcal{R}_{\rm q} = B_{\rm q}/B_{\star}$ and $\mathcal{R}_{\rm o} = B_{\rm o}/B_{\star}$ are the magnetic field ratios where $B_{\star} = B_{\rm d} + B_{\rm q} + B_{\rm o}$. By definition, $\mathcal{R}_{\rm d} + \mathcal{R}_{\rm q} + \mathcal{R}_{\rm o} = 1$. The subscripts $\rm d$, $\rm q$ and $\rm o$ indicate dipole, quadrupole and octopole respectively. Finally, $K_{\rm d} = 1.53$, $K_{\rm q} = 1.7$, $K_{\rm o} = 1.8$, $m_{\rm d} = 0.229$, $m_{\rm q} = 0.134$ and $m_{\rm o} = 0.087$ are fit parameters obtained from the MHD simulations of \citet{Finley2018}. Equation (\ref{eq:BrakingLaw}) has the form of a twice broken power law. Its behaviour is discussed in section 2 of \citet{See2019Dip}.

In the MHD simulations of F18, the prescribed magnetic fields are axisymmetric and potential. This field is specified by the variables $B_{\rm d}$, $B_{\rm q}$ and $B_{\rm o}$ which are the magnetic field strengths at magnetic pole (which is coincident with the rotation pole since the field is axisymmetric) at the stellar surface\footnote{In the language of the spherical harmonic decomposition used to represent the ZDI maps, i.e. equations (2)-(4) in \citet{Lehmann2018ZDI}, the magnetic field configurations at the stellar surface used in FM18 correspond to the $\alpha_{lm}$ and $\beta_{lm}$ terms. These geometries are picked such that the harmonic coefficients satisfy $\alpha_{lm} = \beta_{lm}$ and $\gamma_{lm}=0$ for all $l$ and $m$ values as well as all $m \neq 0$ terms being zero due to the axisymmetry requirement. $B_{\rm d}$, $B_{\rm q}$ and $B_{\rm o}$ are the field strengths of the $l=1$, $l=2$ and $l=3$ components respectively at $r=r_\star$ and $\theta=0$}. Since the maps used by \citet{Lehmann2018ZDI} contain a mixture of axisymmetric and non-axisymmetric modes, we use the surface average unsigned field strengths defined in section \ref{sec:ZDIvsSim} instead of the polar field strengths from the maps when using the F18 braking law, i.e. $B_{\rm d}=\langle B_{\rm sim,d}\rangle$, $B_{\rm q}=\langle B_{\rm sim,q}\rangle$ and $B_{\rm o}=\langle B_{\rm sim,o}\rangle$ when considering the flux transport simulations or $B_{\rm d}=\langle B_{\rm ZDI,d}\rangle$, $B_{\rm q}=\langle B_{\rm ZDI,q}\rangle$ and $B_{\rm o}=\langle B_{\rm ZDI,o}\rangle$ when considering the ZDI reconstructions. By doing this, we have effectively accounted for the non-axisymmetric fields by moving power from the non-axisymmetric modes into the axisymmetric ones. This is justified as \citet{Garraffo2016} has shown that angular momentum loss has a much stronger dependence on magnetic field strength and spherical harmonic degree, $l$ (denoted as $n$ in \citet{Garraffo2016}) than spherical harmonic order, $m$. The alternative, to only consider the axisymmetric modes of the maps from \citet{Lehmann2018ZDI}, would result in underestimated torque estimates since the magnetic flux in the non-axisymmetric modes would not be accounted for \citep[see section 5.1 of][for further discussion on the treatment of non-axisymmetric modes when using braking laws]{Finley2018Sun}. Additionally, we note that using surface averaged field strengths with the F18 braking law will introduce a small systematic error. This is because surface averaged field strengths, $\langle B_{\rm d}\rangle$, $\langle B_{\rm q}\rangle$ and $\langle B_{\rm o}\rangle$, are not equal to polar field strengths $B_{\rm d}$, $B_{\rm q}$ and $B_{\rm o}$ that are required for the F18 braking law even for the axisymmetric geometries considered by F18. However, these errors will mostly cancel out since we calculate ratios of torques rather than absolute torque values in this work (see section \ref{subsec:TorqueRatio}).

As well as the magnetic field strengths, the F18 braking law requires the stellar mass-loss rate in order to calculate a spin-down torque. The advantage of braking laws, such as the F18 braking law, is that they quantify the functional dependence of the torque on the mass-loss rate without requiring any knowledge of how the mass-loss rate depends on parameters such as magnetic field strength or wind-driving mechanisms. Studies that use these types of braking law typically use an independent model to calculate the mass-loss rate \citep[e.g.][]{Gallet2013,Gallet2015,Johnstone2015,Amard2016,See2017,See2018}. However, as we will show in section \ref{subsec:TorqueRatio}, we sidestep the need to calculate a mass-loss rate by calculating a torque ratio instead.
 
\subsection{Torque ratio}
\label{subsec:TorqueRatio}
In this section we will determine how much torques calculated using ZDI maps may be underestimated by. Rather than calculating absolute torque values, we will calculate a ratio of torques, $T_{\rm sim}/T_{\rm ZDI}$, where $T_{\rm sim}$ is the torque calculated using the field strengths from a flux transport simulation and $T_{\rm ZDI}$ is the torque calculated using the field strengths from the corresponding ZDI reconstruction. We will assume that the true mass-loss rate of the star is known, i.e. we will use the same mass-loss rate to calculate $T_{\rm sim}$ and $T_{\rm ZDI}$, without actually specifying what that mass-loss rate is. By doing so, the dependence of the torque ratio on mass-loss rate is removed (see equations (\ref{eq:TorqueRatioDip})-(\ref{eq:TorqueRatioComponents})).

\begin{figure}
	\begin{center}
	\includegraphics[trim=0.5cm 0.5cm 1cm 0cm,width=\columnwidth]{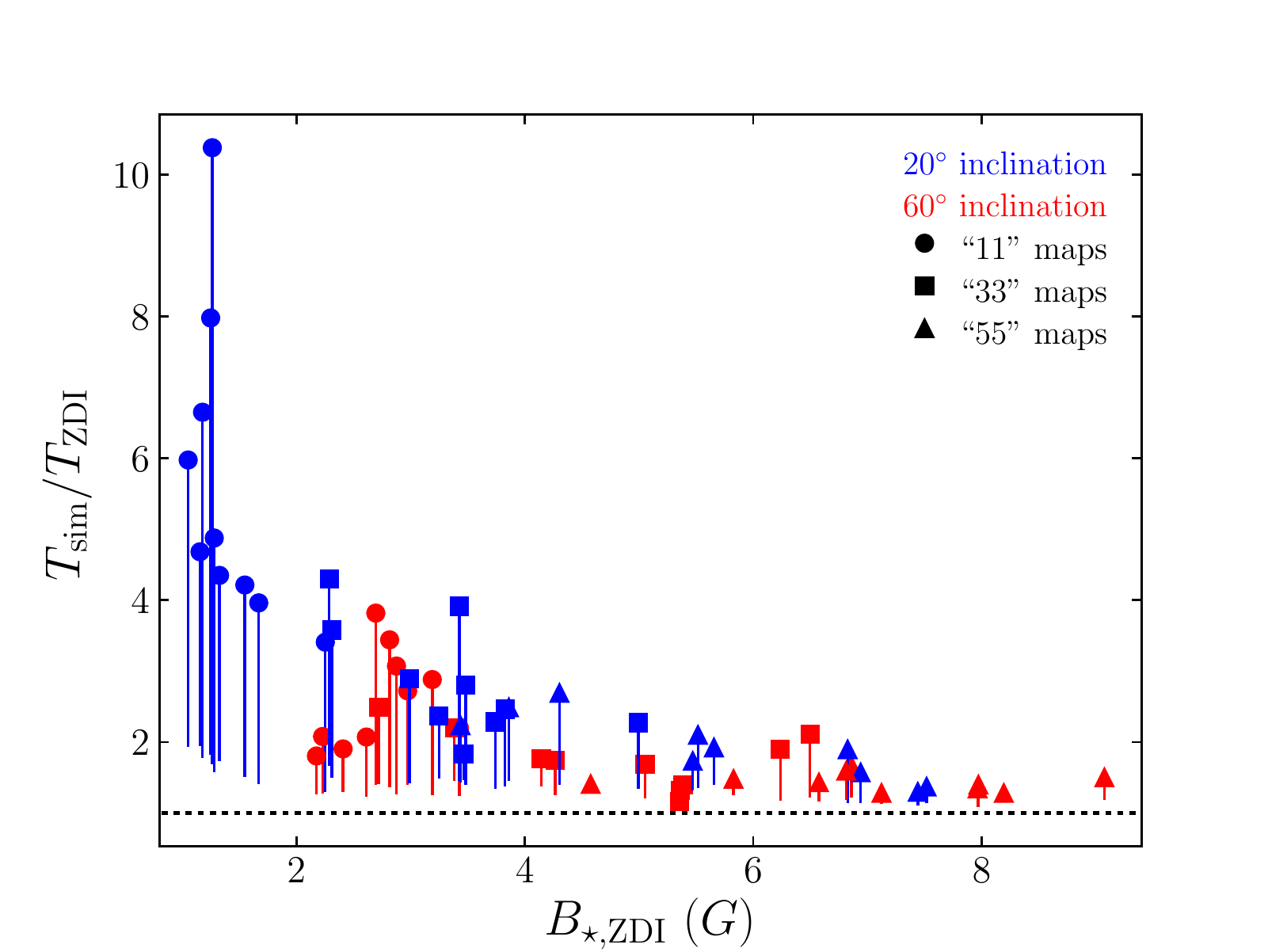}
	\end{center}
	\caption{The amount torques estimated using ZDI maps may be underestimated by, $T_{\rm sim}/T_{\rm ZDI}$, against the field strength reconstructed in the ZDI map. The range of possible $T_{\rm sim}/T_{\rm ZDI}$ values for each ZDI map is shown by a vertical bar. The value of $T_{\rm sim}/T_{\rm ZDI}$, assuming that only the dipole component contributes to the spin-down torque, i.e. equation (\ref{eq:TorqueRatioA}), is plotted on each bar with symbols that have the same meaning as in fig. \ref{fig:BComp}.}
	\label{fig:TorqueComp}
\end{figure}

The F18 braking law has three different regimes, one that depends on the dipole, equation (\ref{eq:BrakingLawDip}), one that depends on the dipole and quadrupole, equation (\ref{eq:BrakingLawQuad}) and one that depends on the dipole, quadrupole and octupole, equation (\ref{eq:BrakingLawOct}). Over the last few years, a number of works have suggested that only the dipole component contributes significantly to the open flux, and hence the spin-down torque, for the majority of stars \citep{See2017,See2018} as well as the Sun \citep{Jardine2017}. We will examine this claim further in section \ref{subsec:MDotCrit} but for now let us assume it is correct. When assuming that the dipole component dominates, the F18 braking law reduces to just equations (\ref{eq:Torque}) and (\ref{eq:BrakingLawDip}). The torque ratio is therefore given by

\begin{equation}
	\frac{T_{\rm sim}}{T_{\rm ZDI}}\bigg\rvert_{\rm d} = \left(\frac{\langle B_{\rm sim,d} \rangle}{\langle B_{\rm ZDI,d} \rangle}\right)^{4m_{\rm d}}.
	\label{eq:TorqueRatioDip}
\end{equation}
The dipole torque ratio only depends on the dipole field strengths from the flux transport simulations and the ZDI reconstructions as well as the power law index in equation (\ref{eq:BrakingLawDip}), $m_{\rm d}$. The dependence on other parameters such as mass-loss rate and rotation period have been normalised out. This means that we do not have to decide on a physical value for these quantities to calculate the torque ratio. We plot the dipole torque ratios in fig. \ref{fig:TorqueComp} as a function of $B_{\star ,{\rm ZDI}}$ using the same symbols as fig. \ref{fig:BComp}. For stars with strong fields, the dipole torque ratio is slightly larger than unity since ZDI does a good job reconstructing the dipole component for these stars (see fig. \ref{fig:BComp}). However, the dipole torque ratio can be as high as $\sim$10 for the stars with the weakest fields in our sample.

Although it is believed that the dipole dominates angular momentum-loss for the majority of low-mass stars, there are some instances when non-dipolar field modes may become important. In these cases, we must consider the full form of equation (\ref{eq:BrakingLaw}) to properly calculate the torque ratio rather than just using equation (\ref{eq:BrakingLawDip}). Calculating $T_{\rm sim}/T_{\rm ZDI}$ is difficult since its value will depend on the mass-loss rate which is an unknown quantity. However, we can determine the range of possible $T_{\rm sim}/T_{\rm ZDI}$ values. This range is independent of mass-loss rate and is given by

\begin{equation}
\begin{split}	
	{\rm min}\left(\frac{T_{\rm sim}}{T_{\rm ZDI}}\right) &={\rm min}\left\{ \frac{T_{\rm sim}}{T_{\rm ZDI}}\bigg\rvert_{\rm d},\frac{T_{\rm sim}}{T_{\rm ZDI}}\bigg\rvert_{\rm q},\frac{T_{\rm sim}}{T_{\rm ZDI}}\bigg\rvert_{\rm o} \right\} \\
	{\rm max}\left(\frac{T_{\rm sim}}{T_{\rm ZDI}}\right) &={\rm max}\left\{ \frac{T_{\rm sim}}{T_{\rm ZDI}}\bigg\rvert_{\rm d},\frac{T_{\rm sim}}{T_{\rm ZDI}}\bigg\rvert_{\rm q},\frac{T_{\rm sim}}{T_{\rm ZDI}}\bigg\rvert_{\rm o} \right\}
\end{split}
\label{eq:TorqueRatio}
\end{equation}
where

\begin{subequations}
\label{eq:TorqueRatioComponents}
\begin{align}
	\frac{T_{\rm sim}}{T_{\rm ZDI}}\bigg\rvert_{\rm d} &= \left(\frac{\langle B_{\rm sim,d} \rangle}{\langle B_{\rm ZDI,d} \rangle}\right)^{4m_{\rm d}}
	\label{eq:TorqueRatioA}\\
	\frac{T_{\rm sim}}{T_{\rm ZDI}}\bigg\rvert_{\rm q} &=  \left(\frac{\langle B_{\rm sim,d} \rangle + \langle B_{\rm sim,q} \rangle}{\langle B_{\rm ZDI,d} \rangle + \langle B_{\rm ZDI,q} \rangle}\right)^{4m_{\rm q}}
	\label{eq:TorqueRatioB}\\
	\frac{T_{\rm sim}}{T_{\rm ZDI}}\bigg\rvert_{\rm o} &= \left(\frac{\langle B_{\rm sim,d} \rangle + \langle B_{\rm sim,q} \rangle + \langle B_{\rm sim,o} \rangle}{\langle B_{\rm ZDI,d} \rangle + \langle B_{\rm ZDI,q} \rangle + \langle B_{\rm ZDI,o} \rangle}\right)^{4m_{\rm o}}.
	\label{eq:TorqueRatioC}
\end{align}
\end{subequations}
Equation (\ref{eq:TorqueRatioA}) is just the dipole torque ratio (equation (\ref{eq:TorqueRatioDip})) which we show again for completeness. Equations (\ref{eq:TorqueRatioB}) and (\ref{eq:TorqueRatioC}) are the equivalent torque ratios that account for the quadrupolar and octupolar components respectively. In fig. \ref{fig:TorqueComp}, we plot the full range of $T_{\rm sim}/T_{\rm ZDI}$ values as a function of $B_{\star ,{\rm ZDI}}$ with vertical bars. The range of possible torque ratio values are shown in table \ref{tab:vals}. In general, we see that the range of possible torque ratio values is larger for stars with weaker magnetic fields. Additionally, the dipole torque ratio tends to be at the upper end of the possible torque ratios. Overall, we find that estimating spin-down torques with ZDI maps becomes more reliable for stars with stronger magnetic fields.

\subsection{The critical mass-loss rate}
\label{subsec:MDotCrit}
In \citet{See2019Dip}, we showed that, for a given magnetic field geometry, high order field modes only contribute to spin-down if the mass-loss rate of the star was sufficiently large. However, if the mass-loss rate was below some critical mass-loss rate, $\dot{M}_{\rm crit}$, only the dipolar field contributes to stellar spin-down. In that work, we concluded that few stars have mass-loss rates that exceed the critical mass-loss rate. However, $\dot{M}_{\rm crit}$ depends on the magnetic map used. In light of the work conducted by \citet{Lehmann2018ZDI}, it is worth re-examining the conclusions of \citet{See2019Dip}.

The critical mass-loss rate of a star is given by

\begin{equation}
	\dot{M}_{\rm crit} = 0.33\frac{B_{\star}^2 r_{\star}^2}{v_{\rm esc}} \frac{\mathcal{R}_{\rm d}^{4.82}}{(\mathcal{R}_{\rm d}+\mathcal{R}_{\rm q})^{2.82}}.
	\label{eq:MDotCrit}
\end{equation}
This is derived by equating equations (\ref{eq:BrakingLawDip}) and (\ref{eq:BrakingLawQuad}) and solving for the mass-loss rate. The numerical values of $K_{\rm d}$, $K_{\rm q}$, $m_{\rm d}$ and $m_{\rm q}$ have been substituted in. This derivation assumes that the first non-dipolar mode that needs to be accounted for is the quadrupolar mode which is true for all the ZDI maps and the majority of the flux transport magnetic field maps we use in this work \citep[see][for further details]{See2019Dip}. Similar to section \ref{subsec:TorqueRatio}, we calculate the ratio of the critical mass-loss rate for each flux transport simulation to the critical mass-loss rate for the associated synthetic ZDI reconstruction. This ratio is given by

\begin{equation}
	\frac{\dot{M}_{\rm crit,sim}}{\dot{M}_{\rm crit,ZDI}} = \left(\frac{\langle B_{\rm sim,d} \rangle}{\langle B_{\rm ZDI,d} \rangle}\right)^{4.82} \left(\frac{\langle B_{\rm ZDI,d} \rangle+\langle B_{\rm ZDI,q} \rangle}{\langle B_{\rm sim,d} \rangle+\langle B_{\rm sim,q} \rangle}\right)^{2.82}.
	\label{eq:MDotCritRatio}
\end{equation}
For eight cases, $\dot{M}_{\rm crit,sim}/\dot{M}_{\rm crit,ZDI}$ is given by equation (\ref{eq:MDotCritRatioAlt}) rather than equation (\ref{eq:MDotCritRatio}). The reason is that the assumptions used to derive equation (\ref{eq:MDotCrit}) do not apply to four of the flux transport magnetic field maps. It turns out that the first non-dipolar mode that must be accounted for in these cases is the octupolar mode. Using equation (\ref{eq:MDotCritRatioAlt}) rather than equation (\ref{eq:MDotCritRatio}) results in only a $\lesssim20\%$ difference. Since this difference is small and only affects a small number of cases, we leave further discussion of these cases to appendix \ref{app:MDotCrit}.

\begin{figure}
	\begin{center}
	\includegraphics[trim=0.5cm 0.5cm 1cm 0cm,width=\columnwidth]{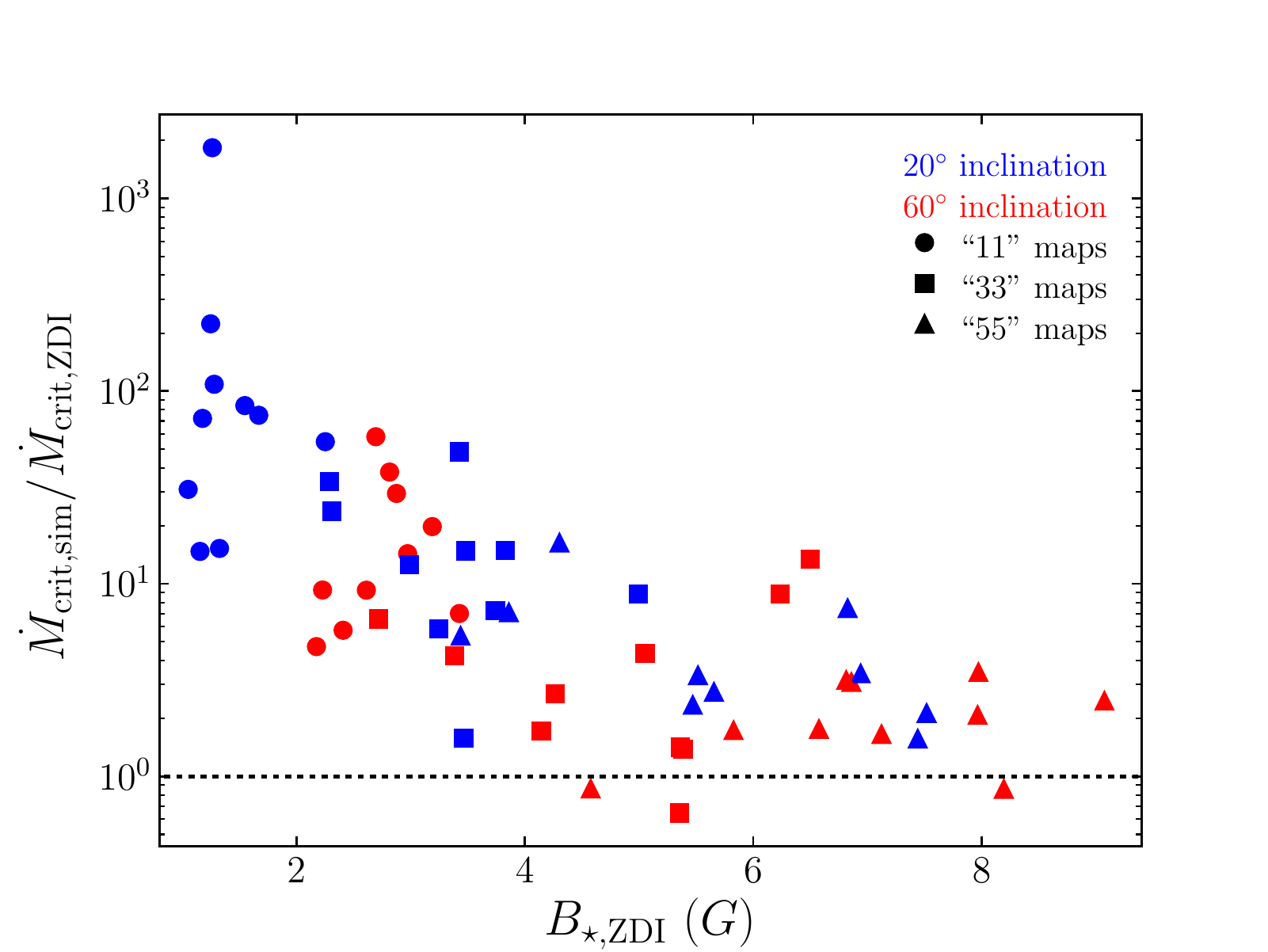}
	\end{center}
	\caption{The amount by which critical mass-loss rates estimated using ZDI maps may be underestimated, $\dot{M}_{\rm crit,sim}/\dot{M}_{\rm crit,ZDI}$, against the field strength reconstructed in the ZDI map. Symbols have the same meaning as in fig. \ref{fig:BComp}.}
	\label{fig:MDotCrit}
\end{figure}

In fig. \ref{fig:MDotCrit}, we plot the ratio of these $\dot{M}_{\rm crit}$ estimates using the same symbols as fig. \ref{fig:BComp}. Their numerical values can be found in table \ref{tab:vals}. We see that the majority of our sample have $\dot{M}_{\rm crit,sim}/\dot{M}_{\rm crit,ZDI}$ values that are greater than 1 with the weak field stars having the highest $\dot{M}_{\rm crit,sim}/\dot{M}_{\rm crit,ZDI}$ values. In \citet{See2019Dip}, we found that the dipole magnetic field does not dominate angular momentum-loss for a small fraction of the stars we studied. These tended to be the stars with large Rossby numbers, i.e. those with weak magnetic fields (see the top row of figure 5 and section 5 of that paper for additional discussion). Since we have shown that $\dot{M}_{\rm crit}$ is underestimated for the majority of simulated stars studied in this work, it is likely that even fewer of the stars in \citet{See2019Dip} will have $\dot{M}>\dot{M}_{\rm crit}$ than indicated in that paper. Overall, we expect the conclusion from \citet{See2019Dip}, that the dipole field mode dominates angular momentum-loss for most low-mass stars, to still hold.

\section{Discussion and conclusions}
\label{sec:Conclusions}
Recently, \citet{Lehmann2018ZDI} showed that Zeeman-Doppler imaging can underestimate the magnetic energy associated with the large-scale magnetic fields of solar-like stars by around an order of magnitude. In this work, we have utilised their results to determine how underestimated angular momentum-loss rate estimates using ZDI maps may be. We find that angular momentum-loss rate estimates for stars with strong fields are relatively accurate but for stars with weak fields, they may be underestimated by a factor of up to $\sim$10. Additionally, we reconfirm previous findings by \citet{See2019Dip} that the dipole component dominates angular momentum-loss for the majority of low-mass stars. \citet{See2019Dip} demonstrated that if the mass-loss rate of a star is below some critical mass-loss rate, its spin-down is dominated by the dipolar component of its magnetic field. Since we have shown that the critical mass-loss rates estimated in that paper are likely to be underestimated, even less stars than indicated by \citet{See2019Dip} will have mass-loss rates that exceed their critical mass-loss rates.

The results we have presented are important to keep in mind when considering angular momentum-loss rates estimates made with ZDI maps and braking laws. For example, \citet{Finley2019} find that their ZDI based angular momentum-loss rate estimates are systematically lower than estimates based on the rotation evolution model of \citet{Matt2015}. Part of the discrepancy may be accounted for by the accuracy of the ZDI technique. Additionally, angular momentum-loss rate estimates from works that directly incorporate ZDI maps into MHD simulations will also be affected by the accuracy of ZDI. Although our results are less quantitatively applicable to these types of studies, the qualitative result that angular momentum-loss rates are underestimated for weak field stars, should still apply.

Lastly, it is important to note some caveats to this work. Firstly, the range of field strengths explored by \citet{Lehmann2018ZDI} is relatively small. The field strengths in the dipolar, quadrupolar and octupolar components of the ZDI maps reconstructed by these authors ranges from $\sim$0.2G to $\sim$4G. The weakest field strengths seen in these components in ZDI maps of real stars is comparable but the strongest field strengths can exceed 1kG (e.g. compare the top row of fig. 2 from \citet{See2019Dip} to the top row of fig. \ref{fig:BComp} in this work). Although fig. \ref{fig:TorqueComp} suggests angular momentum-loss rate estimates should be fairly accurate for stars with stronger field strengths than those explored here, this is based on extrapolating the trends seen in the work of \citet{Lehmann2018ZDI} and so is still not completely certain. Secondly, the simulations used by \citet{Lehmann2018ZDI} used solar flux emergence patterns. This likely means that their results and the results presented here are most applicable to solar-like stars. Other stars with significantly different masses or rotation periods have magnetic field geometries that are different to that of the Sun \citep[e.g.][]{See2015}. A final caveat to address is the toroidal component of stellar magnetic fields that are often not considered in stellar wind models. As such, we have neglected them in this work. However, it is unclear what role the toroidal field recovered in ZDI maps may play in angular momentum-loss \citep[see also][]{Jardine2013}. Further work is required to address these caveats.

\acknowledgments
The authors thank an anonymous referee for useful comments that helped improved the quality of this paper and Gaitee Hussain for useful discussions. VS, SPM and AJF acknowledge funding from the European Research Council (ERC) under the European Unions Horizon 2020 research and innovation programme (grant agreement No 682393 AWESoMeStars). 

\bibliographystyle{yahapj}
\bibliography{ZDIvsSim}

\begin{thebibliography}{}
\providecommand\natexlab[1]{#1}
\providecommand\JournalTitle[1]{#1}

\bibitem[{{Alvarado-G{\'o}mez} {et~al.}(2016){Alvarado-G{\'o}mez}, {Hussain},
  {Cohen}, {Drake}, {Garraffo}, {Grunhut}, \& {Gombosi}}]{AlvaradoGomez2016}
{Alvarado-G{\'o}mez}, J.~D., {Hussain}, G.~A.~J., {Cohen}, O., {et~al.} 2016,
  \href{http://dx.doi.org/10.1051/0004-6361/201628988}{\JournalTitle{A\&A},
  594, A95}

\bibitem[{{Amard} {et~al.}(2016){Amard}, {Palacios}, {Charbonnel}, {Gallet}, \&
  {Bouvier}}]{Amard2016}
{Amard}, L., {Palacios}, A., {Charbonnel}, C., {Gallet}, F., \& {Bouvier}, J.
  2016,
  \href{http://dx.doi.org/10.1051/0004-6361/201527349}{\JournalTitle{A\&A},
  587, A105}

\bibitem[{{Bouvier} {et~al.}(2014){Bouvier}, {Matt}, {Mohanty}, {Scholz},
  {Stassun}, \& {Zanni}}]{Bouvier2014}
{Bouvier}, J., {Matt}, S.~P., {Mohanty}, S., {et~al.} 2014,
  \href{http://dx.doi.org/10.2458/azu_uapress_9780816531240-ch019}{\JournalTitle{Protostars
  and Planets VI}, 433}

\bibitem[{{Brown} {et~al.}(1991){Brown}, {Donati}, {Rees}, \&
  {Semel}}]{Brown1991}
{Brown}, S.~F., {Donati}, J.-F., {Rees}, D.~E., \& {Semel}, M. 1991,
  \JournalTitle{A\&A}, 250, 463

\bibitem[{{Donati} \& {Brown}(1997)}]{Donati1997}
{Donati}, J.-F., \& {Brown}, S.~F. 1997, \JournalTitle{A\&A}, 326, 1135

\bibitem[{{Donati} {et~al.}(2006){Donati}, {Howarth}, {Jardine}, {Petit},
  {Catala}, {Landstreet}, {Bouret}, {Alecian}, {Barnes}, {Forveille},
  {Paletou}, \& {Manset}}]{Donati2006}
{Donati}, J.-F., {Howarth}, I.~D., {Jardine}, M.~M., {et~al.} 2006,
  \href{http://dx.doi.org/10.1111/j.1365-2966.2006.10558.x}{\JournalTitle{MNRAS},
  370, 629}

\bibitem[{{Finley} \& {Matt}(2017)}]{Finley2017}
{Finley}, A.~J., \& {Matt}, S.~P. 2017,
  \href{http://dx.doi.org/10.3847/1538-4357/aa7fb9}{\JournalTitle{ApJ}, 845,
  46}

\bibitem[{{Finley} \& {Matt}(2018)}]{Finley2018}
---. 2018,
  \href{http://dx.doi.org/10.3847/1538-4357/aaaab5}{\JournalTitle{ApJ}, 854,
  78}

\bibitem[{{Finley} {et~al.}(2018){Finley}, {Matt}, \& {See}}]{Finley2018Sun}
{Finley}, A.~J., {Matt}, S.~P., \& {See}, V. 2018,
  \href{http://dx.doi.org/10.3847/1538-4357/aad7b6}{\JournalTitle{ApJ}, 864,
  125}

\bibitem[{{Finley} {et~al.}(2019){Finley}, {See}, \& {Matt}}]{Finley2019}
{Finley}, A.~J., {See}, V., \& {Matt}, S.~P. 2019,
  \href{http://dx.doi.org/10.3847/1538-4357/ab12d2}{\JournalTitle{ApJ}, 876,
  44}

\bibitem[{{Gallet} \& {Bouvier}(2013)}]{Gallet2013}
{Gallet}, F., \& {Bouvier}, J. 2013,
  \href{http://dx.doi.org/10.1051/0004-6361/201321302}{\JournalTitle{A\&A},
  556, A36}

\bibitem[{{Gallet} \& {Bouvier}(2015)}]{Gallet2015}
---. 2015,
  \href{http://dx.doi.org/10.1051/0004-6361/201525660}{\JournalTitle{A\&A},
  577, A98}

\bibitem[{{Garraffo} {et~al.}(2016){Garraffo}, {Drake}, \&
  {Cohen}}]{Garraffo2016}
{Garraffo}, C., {Drake}, J.~J., \& {Cohen}, O. 2016,
  \href{http://dx.doi.org/10.1051/0004-6361/201628367}{\JournalTitle{A\&A},
  595, A110}

\bibitem[{{Gibb} {et~al.}(2016){Gibb}, {Mackay}, {Jardine}, \&
  {Yeates}}]{Gibb2016}
{Gibb}, G.~P.~S., {Mackay}, D.~H., {Jardine}, M.~M., \& {Yeates}, A.~R. 2016,
  \href{http://dx.doi.org/10.1093/mnras/stv2920}{\JournalTitle{MNRAS}, 456,
  3624}

\bibitem[{{Hussain} {et~al.}(2016){Hussain}, {Alvarado-G{\'o}mez}, {Grunhut},
  {Donati}, {Alecian}, {Oksala}, {Morin}, {Fares}, {Jardine}, {Drake}, {Cohen},
  {Matt}, {Petit}, {Redfield}, \& {Walter}}]{Hussain2016}
{Hussain}, G.~A.~J., {Alvarado-G{\'o}mez}, J.~D., {Grunhut}, J., {et~al.} 2016,
  \href{http://dx.doi.org/10.1051/0004-6361/201526595}{\JournalTitle{A\&A},
  585, A77}

\bibitem[{{Jardine} {et~al.}(2017){Jardine}, {Vidotto}, \& {See}}]{Jardine2017}
{Jardine}, M., {Vidotto}, A.~A., \& {See}, V. 2017,
  \href{http://dx.doi.org/10.1093/mnrasl/slw206}{\JournalTitle{MNRAS}, 465,
  L25}

\bibitem[{{Jardine} {et~al.}(2013){Jardine}, {Vidotto}, {van Ballegooijen},
  {Donati}, {Morin}, {Fares}, \& {Gombosi}}]{Jardine2013}
{Jardine}, M., {Vidotto}, A.~A., {van Ballegooijen}, A., {et~al.} 2013,
  \href{http://dx.doi.org/10.1093/mnras/stt181}{\JournalTitle{MNRAS}, 431, 528}

\bibitem[{{Johnstone} {et~al.}(2015){Johnstone}, {G{\"u}del}, {Brott}, \&
  {L{\"u}ftinger}}]{Johnstone2015}
{Johnstone}, C.~P., {G{\"u}del}, M., {Brott}, I., \& {L{\"u}ftinger}, T. 2015,
  \href{http://dx.doi.org/10.1051/0004-6361/201425301}{\JournalTitle{A\&A},
  577, A28}

\bibitem[{{Kochukhov} \& {Shulyak}(2019)}]{Kochukhov2019}
{Kochukhov}, O., \& {Shulyak}, D. 2019,
  \href{http://dx.doi.org/10.3847/1538-4357/ab06c5}{\JournalTitle{\apj}, 873,
  69}

\bibitem[{{Lehmann} {et~al.}(2018){Lehmann}, {Hussain}, {Jardine}, {Mackay}, \&
  {Vidotto}}]{Lehmann2018ZDI}
{Lehmann}, L.~T., {Hussain}, G.~A.~J., {Jardine}, M.~M., {Mackay}, D.~H., \&
  {Vidotto}, A.~A. 2018, \JournalTitle{ArXiv e-prints},
  \href{http://arxiv.org/abs/1811.03703}{{\sffamily arXiv:1811.03703
  [astro-ph.SR]}}

\bibitem[{{Matt} {et~al.}(2015){Matt}, {Brun}, {Baraffe}, {Bouvier}, \&
  {Chabrier}}]{Matt2015}
{Matt}, S.~P., {Brun}, A.~S., {Baraffe}, I., {Bouvier}, J., \& {Chabrier}, G.
  2015,
  \href{http://dx.doi.org/10.1088/2041-8205/799/2/L23}{\JournalTitle{ApJl},
  799, L23}

\bibitem[{{Matt} {et~al.}(2012){Matt}, {MacGregor}, {Pinsonneault}, \&
  {Greene}}]{Matt2012}
{Matt}, S.~P., {MacGregor}, K.~B., {Pinsonneault}, M.~H., \& {Greene}, T.~P.
  2012,
  \href{http://dx.doi.org/10.1088/2041-8205/754/2/L26}{\JournalTitle{ApJL},
  754, L26}

\bibitem[{{Morin} {et~al.}(2010){Morin}, {Donati}, {Petit}, {Delfosse},
  {Forveille}, \& {Jardine}}]{Morin2010}
{Morin}, J., {Donati}, J.-F., {Petit}, P., {et~al.} 2010,
  \href{http://dx.doi.org/10.1111/j.1365-2966.2010.17101.x}{\JournalTitle{MNRAS},
  407, 2269}

\bibitem[{{Nicholson} {et~al.}(2016){Nicholson}, {Vidotto}, {Mengel},
  {Brookshaw}, {Carter}, {Petit}, {Marsden}, {Jeffers}, {Fares}, \& {BCool
  Collaboration}}]{Nicholson2016}
{Nicholson}, B.~A., {Vidotto}, A.~A., {Mengel}, M., {et~al.} 2016,
  \href{http://dx.doi.org/10.1093/mnras/stw731}{\JournalTitle{MNRAS}, 459,
  1907}

\bibitem[{{Pantolmos} \& {Matt}(2017)}]{Pantolmos2017}
{Pantolmos}, G., \& {Matt}, S.~P. 2017,
  \href{http://dx.doi.org/10.3847/1538-4357/aa9061}{\JournalTitle{ApJ}, 849,
  83}

\bibitem[{{Reiners} \& {Basri}(2009)}]{Reiners2009}
{Reiners}, A., \& {Basri}, G. 2009,
  \href{http://dx.doi.org/10.1051/0004-6361:200811450}{\JournalTitle{A\&A},
  496, 787}

\bibitem[{{R{\'e}ville} {et~al.}(2015){R{\'e}ville}, {Brun}, {Matt},
  {Strugarek}, \& {Pinto}}]{Reville2015}
{R{\'e}ville}, V., {Brun}, A.~S., {Matt}, S.~P., {Strugarek}, A., \& {Pinto},
  R.~F. 2015,
  \href{http://dx.doi.org/10.1088/0004-637X/798/2/116}{\JournalTitle{ApJ}, 798,
  116}

\bibitem[{{R{\'e}ville} {et~al.}(2016){R{\'e}ville}, {Folsom}, {Strugarek}, \&
  {Brun}}]{Reville2016}
{R{\'e}ville}, V., {Folsom}, C.~P., {Strugarek}, A., \& {Brun}, A.~S. 2016,
  \href{http://dx.doi.org/10.3847/0004-637X/832/2/145}{\JournalTitle{ApJ}, 832,
  145}

\bibitem[{{Ros{\'e}n} {et~al.}(2015){Ros{\'e}n}, {Kochukhov}, \&
  {Wade}}]{Rosen2015}
{Ros{\'e}n}, L., {Kochukhov}, O., \& {Wade}, G.~A. 2015,
  \href{http://dx.doi.org/10.1088/0004-637X/805/2/169}{\JournalTitle{ApJ}, 805,
  169}

\bibitem[{{See} {et~al.}(2015){See}, {Jardine}, {Vidotto}, {Donati}, {Folsom},
  {Boro Saikia}, {Bouvier}, {Fares}, {Gregory}, {Hussain}, {Jeffers},
  {Marsden}, {Morin}, {Moutou}, {do Nascimento}, {Petit}, {Ros{\'e}n}, \&
  {Waite}}]{See2015}
{See}, V., {Jardine}, M., {Vidotto}, A.~A., {et~al.} 2015,
  \href{http://dx.doi.org/10.1093/mnras/stv1925}{\JournalTitle{MNRAS}, 453,
  4301}

\bibitem[{{See} {et~al.}(2017){See}, {Jardine}, {Vidotto}, {Donati}, {Boro
  Saikia}, {Fares}, {Folsom}, {H{\'e}brard}, {Jeffers}, {Marsden}, {Morin},
  {Petit}, {Waite}, \& {BCool Collaboration}}]{See2017}
---. 2017, \href{http://dx.doi.org/10.1093/mnras/stw3094}{\JournalTitle{MNRAS},
  466, 1542}

\bibitem[{{See} {et~al.}(2018){See}, {Jardine}, {Vidotto}, {Donati}, {Boro
  Saikia}, {Fares}, {Folsom}, {Jeffers}, {Marsden}, {Morin}, {Petit}, \& {BCool
  Collaboration}}]{See2018}
---. 2018, \href{http://dx.doi.org/10.1093/mnras/stx2599}{\JournalTitle{MNRAS},
  474, 536}

\bibitem[{{See} {et~al.}(2019{\natexlab{a}}){See}, {Matt}, {Finley}, {Folsom},
  {Boro Saikia}, {Donati}, {Fares}, {H{\'e}brard}, {Jardine}, {Jeffers},
  {Marsden}, {Mengel}, {Morin}, {Petit}, {Vidotto}, {Waite}, \& {the BCool
  Collaboration}}]{See2019Dip}
{See}, V., {Matt}, S.~P., {Finley}, A.~J., {et~al.} 2019{\natexlab{a}},
  \href{http://dx.doi.org/10.3847/1538-4357/ab46b2}{\JournalTitle{ApJ}, 886,
  120}

\bibitem[{{See} {et~al.}(2019{\natexlab{b}}){See}, {Matt}, {Folsom}, {Boro
  Saikia}, {Donati}, {Fares}, {Finley}, {H{\'e}brard}, {Jardine}, {Jeffers},
  {Lehmann}, {Marsden}, {Mengel}, {Morin}, {Petit}, {Vidotto}, {Waite}, \& {The
  BCool Collaboration}}]{See2019}
{See}, V., {Matt}, S.~P., {Folsom}, C.~P., {et~al.} 2019{\natexlab{b}},
  \href{http://dx.doi.org/10.3847/1538-4357/ab1096}{\JournalTitle{ApJ}, 876,
  118}

\bibitem[{{Semel}(1989)}]{Semel1989}
{Semel}, M. 1989, \JournalTitle{A\&A}, 225, 456

\bibitem[{{Vidotto} {et~al.}(2015){Vidotto}, {Fares}, {Jardine}, {Moutou}, \&
  {Donati}}]{Vidotto2015}
{Vidotto}, A.~A., {Fares}, R., {Jardine}, M., {Moutou}, C., \& {Donati}, J.-F.
  2015, \href{http://dx.doi.org/10.1093/mnras/stv618}{\JournalTitle{MNRAS},
  449, 4117}

\bibitem[{{Vidotto} {et~al.}(2014){Vidotto}, {Jardine}, {Morin}, {Donati},
  {Opher}, \& {Gombosi}}]{Vidotto2014Torque}
{Vidotto}, A.~A., {Jardine}, M., {Morin}, J., {et~al.} 2014,
  \href{http://dx.doi.org/10.1093/mnras/stt2265}{\JournalTitle{MNRAS}, 438,
  1162}

\end{thebibliography}

\appendix
\section{Critical mass-loss rate calculation}
\label{app:MDotCrit}
In section \ref{subsec:MDotCrit}, we calculated the ratio of critical mass-loss rates using equation (\ref{eq:MDotCritRatio}). However, for a small number of cases, equation (\ref{eq:MDotCritRatio}) does not apply. In this appendix, we show the correct form for the ratio of critical mass-loss rates for these cases. It was stated that the critical mass-loss rate is given by equation (\ref{eq:MDotCrit}) which is found by equating equations (\ref{eq:BrakingLawDip}) and (\ref{eq:BrakingLawQuad}) and solving for the mass-loss rate. This expression is derived under the assumption that, for a sufficiently high mass-loss rate, the first non-dipolar field mode to contribute to angular momentum-loss is the quadrupolar field component. This turns out to be true for all the ZDI reconstructions we have analysed as well as the majority of the flux transport simulations. However, the assumption does not hold for 4 of the flux transport simulations. In these cases, the first non-dipolar field mode to contribute to angular momentum-loss is the octupolar field component. The critical mass-loss rate for these cases is given by equating equations (\ref{eq:BrakingLawDip}) and (\ref{eq:BrakingLawOct}) and solving for the mass-loss rate to give

\begin{equation}
	\dot{M}_{\rm crit} = 0.32\frac{B_{\star}^2 r_{\star}^2}{v_{\rm esc}} \frac{\mathcal{R}_{\rm d}^{3.23}}{(\mathcal{R}_{\rm d}+\mathcal{R}_{\rm q}+\mathcal{R}_{\rm o})^{1.23}}.
	\label{eq:MDotCritAlt}
\end{equation}
For these 4 cases, equation (\ref{eq:MDotCritAlt}) gives a smaller value for $\dot{M}_{\rm crit}$ than equation (\ref{eq:MDotCrit}). Consequently, the ratio of critical mass-loss rates has a different form to that given by equation (\ref{eq:MDotCritRatio}). Using equation (\ref{eq:MDotCritAlt}) to calculate the critical mass-loss rate for the flux transport simulations, $\dot{M}_{\rm crit,sim}$, and recalling that the critical mass-loss rate for the corresponding ZDI reconstructions, $\dot{M}_{\rm crit,ZDI}$, is still given by equation (\ref{eq:MDotCrit}), the ratio of critical mass-loss rates is given by 

\begin{equation}
	\frac{\dot{M}_{\rm crit,sim}}{\dot{M}_{\rm crit,ZDI}} = 0.97 \frac{\langle B_{\rm sim,d}\rangle^{3.23}}{\left(\langle B_{\rm sim,d}\rangle+\langle B_{\rm sim,q}\rangle+\langle B_{\rm sim,o}\rangle\right)^{1.23}} \  \frac{\left(\langle B_{\rm ZDI,d}\rangle+\langle B_{\rm ZDI,q}\rangle\right)^{2.82}}{\langle B_{\rm ZDI,d}\rangle^{4.82}}.
	\label{eq:MDotCritRatioAlt}
\end{equation}
Using equation (\ref{eq:MDotCritRatioAlt}) rather than equation (\ref{eq:MDotCritRatio}) results in a change of $\sim 20\%$ at most for these 4 cases.

\end{document}